# Enhanced Ferroelectricity and Spin Current Waves in M-Type Barium Hexaferrite

*Xue Li, Nan Nan, Guo-Long Tan*[*]

*State Key Laboratory of Advanced Technology for Materials Synthesis and Processing,*

*Wuhan University of Technology, Wuhan 430070, China*

**Abstract**

The intrinsic ferroelectricity and related dielectric properties of M-type Barium Hexaferrite ($BaFe_{12}O_{19}$) with excellent magnetic performance are reported in this paper. A classic electric polarization (P-E) hysteresis loop with full saturation, together with two nonlinear reversal current peaks in the I-V curve and huge change of dielectric constant in the vicinity of Curie temperature, have all demonstrated themselves as sufficient experimental evidences to verify the ferroelectricity of $BaFe_{12}O_{19}$ ceramics. It holds a large remnant polarization at 108 μC/cm² and a suitable coercive field at 117 kV/cm. Two peaks at 194°C and 451°C in the temperature-dependent dielectric spectrum of $BaFe_{12}O_{19}$ ceramics are considered to be the phase transition from ferro- to antiferro- and antiferro- to para-electric structures. A conventional strong magnetic hysteresis loop was also observed. The magnetically induced polarization upon the $BaFe_{12}O_{19}$ ceramics was achieved in the form of alternating spin current waves. A ME-coupling voltage with an amplitude of ±23mV on an applied magnetic field at 500mT was achieved. These combined multiple functional responses of large ferroelectrics and strong ferromagnetism reveal the excellent multiferroic features of $BaFe_{12}O_{19}$, which would bring forth the great opportunity to create novel electric devices with active coupling effect between magnetic and electric orders.

## 1. Introduction

The traditional information storage and sensors required a large and energy consuming electromagnet to control the direction of magnetization and regulate the magnetoresistance. Those conventional materials possessed the promising physical effect for favoring functions, like anisotropic magnetoresistance (AMR) [1] and giant magnetoresistance (GMR) [2], usually cause big energy consumption to modulate the direction of magnetization and resistance from the indirect magnetic field raised by big current. It is different from the demand for faster, smaller and ultra-low power to electronic devices in future [3]. So recently the electric field straightforward changing the magnetic properties would be pushed to appear in front of us [4-8]. It sets a new standard of the next generation of spintronic devices over passive magnetoelectronic components which combine memory and logic functions together. Therefore, new multi-functional materials, whose properties can be controlled by several uncorrelated stimulus affecting different order parameters, is highly desired to control the freedom degree of spin and magnetic states by current [9]. The change in ARM and GMR devices with high energy consumption results in the possible generation of novel more energy-saving tunable electronic or spintronic devices [4, 10-12]. Alternatively, multiferroics (MF), exhibiting ferroelectricity and ferromagnetism simultaneously, have the ability to change the magnetic state by applying an E-field through magnetoelectric (ME) coupling effect [13, 14] and could provide the opportunity to manipulate the multiple response of the structures by different stimuli sources.

[*] Corresponding author; Tel: +86-27-87870271; fax: +86-27-87879468. Email address: gltan@whut.edu.cn

MF materials exhibit two or more ferroic orders.[15] The magneto-electric coupling effect in MF materials can be used for the design of more advanced memory devices, which can be written in information by electric field and read out magnetically. This way can apply electric field to alter the storage units and avoid the information changes in reading out [16]. Considering the root reasons of ME coupling, the single phase MF materials may fully reflect this effect. The vision of this function can solve the miniaturization problem of magnetic random access memory (MRAM) written by magnetic fields or large currents [7]. Therefore, it's meaningful to expand the development of diverse MF materials.

Besides those extensively researched MF compounds, such as $BiFeO_3$[5], $TbMnO_3$[17] and $DyMnO_3$[18], a new kind of ferrite which termed hexaferrites with hexagonal structure, divided into M-type, Y-type and Z-type, draw attention for the obvious ME effects and highly sensitive to magnetic field [19, 21]. However, the ferroelectricity of these ferrites was not reported, their ferromagnetism was not strong enough for practical application. The milestone in multiferroics would be the realization of simultaneous coexistence of large ferroelectricity and strong ferromagnetism, together with giant ME coupling effects in one single phase at room temperature. Hence, it is a long standing challenge in seeking multiferroics with practicable ferroelectric and magnetic performance [23] as well as excellent ME sensitivity.[24]

M-type lead hexaferrite, a traditional ferrimagnetic compound, was recently revealed to be ferroelectrics.[25, 26] These M-type hexaferrites not only demonstrate excellent magnetic properties,[27] such as high saturation magnetization, large coercivity and corrosion resistivity, [28] but also exhibit large ferroelectricity.[25, 26] It may be the candidate which would be able to step across the multiferroic milestone. Most recently, $BaFe_{12}O_{19}$ single crystal were claimed to be antiferroelectric phase and present some quantum spin liquid frustration at a temperature of 2K. [29, 30] Meanwhile, doubtful ferroelectricity of $BaFe_{12}O_{19}$ ceramics and crystals had already been reported in several research groups a few years ago. [29-32] However, those published electric hysteresis loops are banana shaped and were not saturated due to large current leakage and are unsimilar to the previous research result of traditional ferroelectric materials [29-32]. The source of current leakage is consequent from the high concentration of oxygen vacancies and $Fe^{2+}$ ions inside the bulk specimens. Therefore, the ferroelectricity of $BaFe_{12}O_{19}$ ceramics still remains controversial and uncertain, the unsaturated hysteresis loops are not powerful evidences [29-32]. In order to further confirm the ferroelectricity of this material, the synthesis process of the $BaFe_{12}O_{19}$ ceramics was meliorated by replacing annealing atmosphere from air to oxygen which aims to reduce the oxygen vacancies and the possibility of the chance of Fe atom value state. In this way, the current leakage would be massively decreased and makes the appearance of saturated P-E hysteresis loop become possible, as it appeared in the multiferroic $PbFe_{12}O_{19}$ and $SrFe_{12}O_{19}$ specimens before [26, 33, 34]. In this paper, we will exhibit the improved ferroelectric performance, enhanced impedance property, large dielectric anomaly near the Curie temperature, nonlinear I-V peaks, remarkable ME response, together with strong ferromagnetism in the M-type Barium hexaferrite ($BaFe_{12}O_{19}$) ceramics, which is lead-free and environment-friendly. Experimental evidences for proving the ferroelectricity of $BaFe_{12}O_{19}$ ceramics will be provided and discussed.

## 2. Experimental Procedure

A polymer precursor procedure was used to synthesize $BaFe_{12}O_{19}$ powders for accurately controlling element atomic ratio in the target compound. The raw materials include barium and iron source which are Barium acetate $Ba(CH_3COO)_2 \cdot 3H_2O)$ (99.9%, Aladin) and ferric acetylacetonate ($C_{15}H_{21}FeO_6$) (99.9%, Alfa Aesar). The solvents include ethanol, acetone and glycerin. At beginning, 0.60g iron acetylacetone was weighed and dissolved in a mixture solvent of 50ml ethanol and 70ml acetone in a 250ml three-flask inside a glove box. The mixture solution was maintained on the mantle for heating and stirring at 70°C for 6 hours. While outside the glove box, barium acetate was weighed by an atomic ratio of Ba : Fe in 1:9.5~1:10 and then dissolved in glycerin in a 50 mL flask. To avoid moisture, the barium dispersion solution was distilled at 120 °C for 1 hour in a rotary evaporator. After full dissolution, the barium precursor solution was transported into the same glove box, where two precursor iron and barium solutions were mixed together and maintained on the mantle at 70°C for 6 hours. Afterwards, 45 mL of ammonia was added into the mixture precursor solution, which was continually heated and stirred on the mantle at 70°C for more than 24 hours, so as to precipitate Ba and Fe ions completely. Finally, the synthesized particles suspended in the solution were separated from the solvent by centrifugation and then calcined at 500°C and 800°C for 1 hour each to ensure the removal of organic impurities. After calcination and grinding, pure barium hexaferrite ($BaFe_{12}O_{19}$) nano-powders in single-phase were obtained. The bulk ceramic samples were prepared by pressing the powders into a circular mould with an inner diameter of 0.6mm, sintering the pressed pellet at 1250°C for 2 hours. Now we come to the most important step, in which the powder and pellet ceramic samples were annealed in a tube furnace with pure oxygen atmosphere. The specific operation is annealing the sample at 700°C in pure $O_2$ for 3 hours, turning it over up-side-down and annealing it again for another 3 hours, then heating it at 600°C in pure oxygen atmosphere for last 3 hours in a tube furnace. Both sides of the pellet ceramics shall be exposed to pure $O_2$ during heat treatment process. In this way, powders, bulk pellets and capacitor (the bulk pellet being coated with silver electrode) samples were successfully fabricated. Phase analysis of $BaFe_{12}O_{19}$ powders and bulk pellet samples was performed by X-ray diffraction (XRD) with Cu–$K_\alpha$ radiation. Magnetization was measured by a Quantum Design physical property measurement system (PPMS). The P-E hysteresis loop was measured on a Sawyer-Tower circuit based ferroelectric measurement system. The temperature-dependent dielectric property and impedance spectrum were measured on an LCR instrument (IM3533-01) at the frequency range from 0.01Hz to 200kHz. The magnetoelectric polarization parameters of the capacitor were measured by a Keysight 2450 source meter applying a variable DC magnetic field.

## 3. Results and Discussion

### 3.1 Structure identification of $BaFe_{12}O_{19}$ Ceramics

The structure of the as-prepared $BaFe_{12}O_{19}$ powders was identified by an X-Ray diffraction spectrometer. Both powder specimens and ceramics were fabricated by sintering the specimen at 1250°C for 2 hours and subsequently annealing it in $O_2$ for total period of 9 hours in three stepwise. Figure 1 exhibits the X-Ray diffraction (XRD) spectrum of the as-fabricated $BaFe_{12}O_{19}$ powders, the separated red lines are derived from standard diffraction card of $BaFe_{12}O_{19}$ (PDF#*43-0002*). Obviously, all the specimen diffraction peaks are in good agreement with ones in the standard card. No oxides or other ferrite impurities were found, indicating that pure $BaFe_{12}O_{19}$ powders in single magnetoplumbite-5H structure have been successfully fabricated.

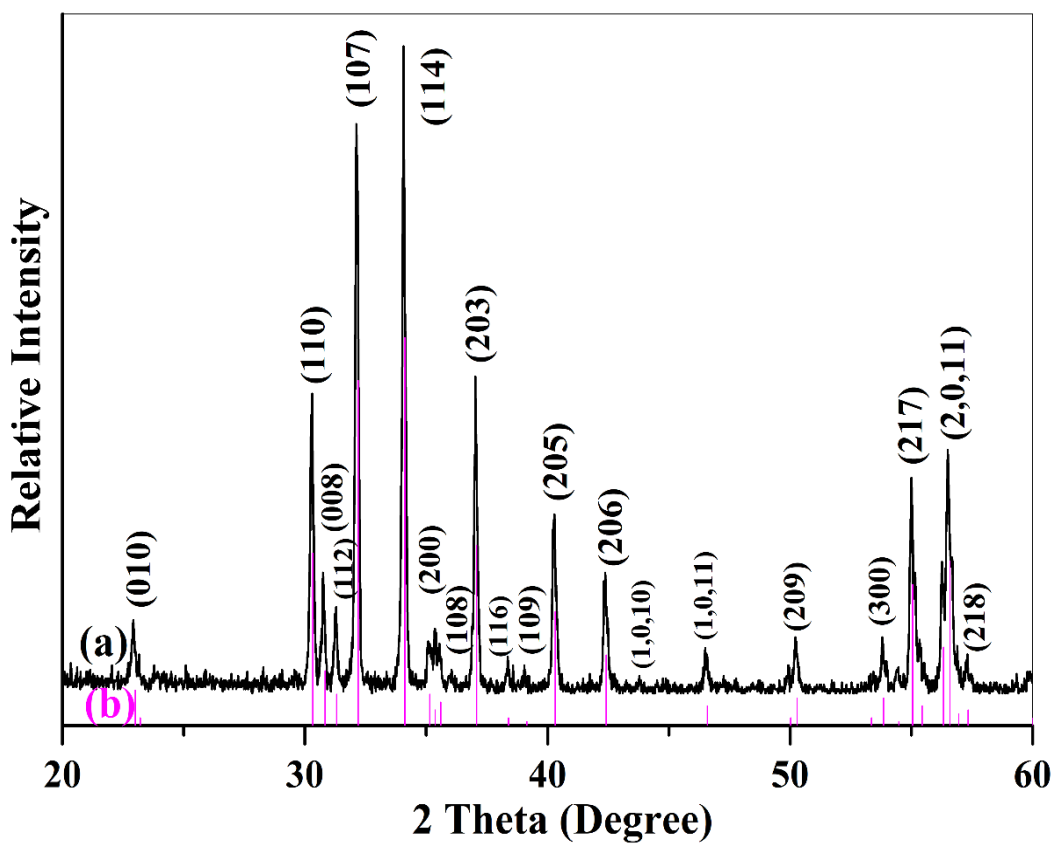

Figure 1: *XRD spectrum of $BaFe_{12}O_{19}$ powders which were calcined at 1250 °C for 2 hours and annealed at 700 °C in $O_2$ for 9hours (a) and Standard XRD spectrum of the $BaFe_{12}O_{19}$ (PDF#43-0002) from discrete red lines (b).*

### 3.2 Impedance Spectrum of $BaFe_{12}O_{19}$ Ceramics

The complex impedance spectrum of the $BaFe_{12}O_{19}$ capacitors was measured upon a Hioki IM3533-01 LCR meter to evaluate the role of oxygen treatment on the electric properties of $BaFe_{12}O_{19}$. *Figure 2*A(a) presents two impedance spectra of $BaFe_{12}O_{19}$ capacitors, one for the specimen being sintered in air only and the other one for the specimen with subsequent $O_2$ annealing process after sintering. Both spectra are composed of only one Cole circle. The diameter of the Cole circle for air-sintered capacitor in *Figure 2*a is around $1.3 \times 10^5 \Omega$, while the other one for $O_2$ annealed capacitor in *Figure 2*A(b) reaches as high as $2.5 \times 10^7 \Omega$. These Cole circles could be expressed as an equivalent circuit in which a capacitor and a resistor are parallelly connected. They reflect the contribution of grains to the electric conduction only, implying that the grain boundaries may make very little contribution to the electric conduction of the $BaFe_{12}O_{19}$ capacitors, which is similar to $PbFe_{12}O_{19}$[26] but different from $SrFe_{12}O_{19}$.[33] Obviously, the diameter of the Cole circle in the $O_2$ annealed specimen is much larger than that in air-sintered specimen, suggesting the oxygen treated $BaFe_{12}O_{19}$ specimen has much higher resistivity than the one without $O_2$ heat treatment.

In order to examine how the oxygen treatment would make impact on the electric properties of $BaFe_{12}O_{19}$ ceramics and how the impedance modules change with the content of $O^{2-}$ vacancies and $Fe^{2+}$, we then calculated the complex impedance modules of the two types of specimens for comparison. The complex impedance of $BaFe_{12}O_{19}$ ceramics is expressed as follows:

$$Z = Z' + jZ'' = \frac{R}{1+(\omega RC)^2} - j\frac{\omega R^2 C}{1+(\omega RC)^2} \qquad \text{(Equation 1)}$$

While the module of the complex impedance may be written as:

$$|Z| = \sqrt{Z'^2 + Z''^2} \qquad \text{(Equation 2)}$$

The impedance modules are corresponding to the dimension of resistivity, which may give expression to content level of $O^{2-}$ vacancies and $Fe^{2+}$ in $BaFe_{12}O_{19}$ ceramics. Larger impedance module implies lower content of the charge carriers in form of $O^{2-}$ vacancies and $Fe^{2+}$, which could greatly reduce the current leakage. The defects of these oxygen vacancies would form a dopant energy level just beneath the bottom of the conduction band in $BaFe_{12}O_{19}$. The charge carriers coming from the excess electrons on oxygen vacancies as well as the electrons hopping across $Fe^{2+}$ to $Fe^{3+}$ could then occupy the defect levels. The ceramics were coated with silver electrodes on the surfaces, forming a capacitor like device. The interface between the electrode and the ceramic surface shapes up to a profile of Schottky barrier, whose potential is usually less than 3V. Upon the electric measurement, the applied electric field, usually is much larger than the potential height of the barrier, would then drive the charge carriers on the defect level to pass over the Schottky barrier at the interface and flow into the silver electrode, as such leakage current flow appears. Based on the Schottky barrier model, the leakage current is proportional to the concentration of the charge carriers on the defect level, suggesting that the higher concentration of the oxygen vacancies and $Fe^{2+}$ would produce larger leakage current.

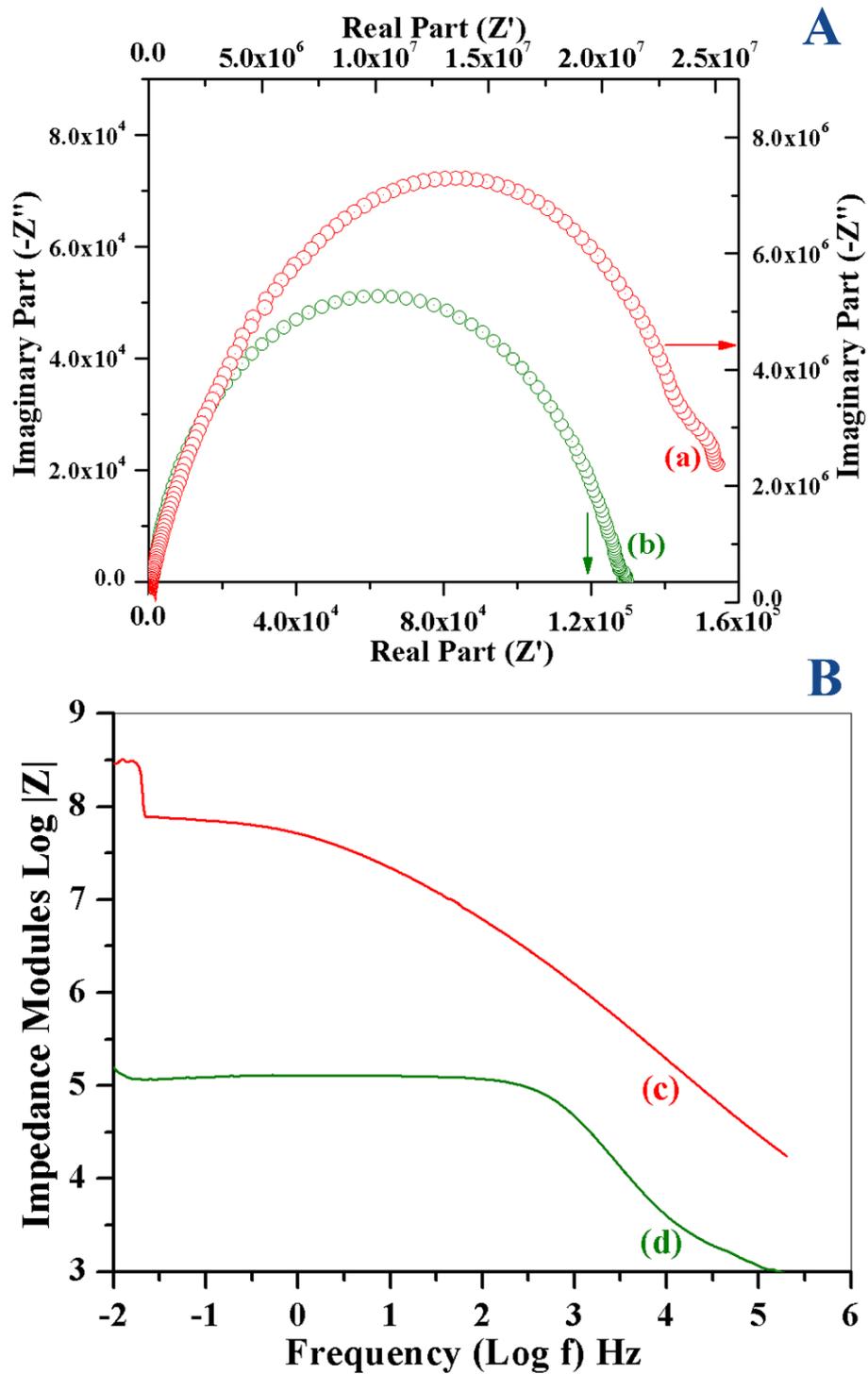

*Figure 2 (A): Complex impedance spectra of the BaFe$_{12}$O$_{19}$ ceramics varying with the frequency from 0.01 Hz to 200 kHz, (a) the cole-cole circle for BaFe$_{12}$O$_{19}$ ceramic with subsequent O$_2$ treatment at 700°C in three step wise; (b) the cole-cole circle for the same specimen without O$_2$ heat treatment. (B): Modules of the complex impedance for (c) the BaFe$_{12}$O$_{19}$ ceramics without O$_2$ heat treatment, (d) the O$_2$ treated ceramic after sintering.*

Under this consideration, the O$_2$ treatment process, which was designed to remove large amount of oxygen vacancies and transform Fe$^{2+}$ into Fe$^{3+}$ for the enhancement of the resistivity, would then greatly reduce the concentration of those charge carriers and thus greatly decrease the current leakage. *Figure 2*B shows the frequency-dependent impedance module or resistivity of these two types of BaFe$_{12}$O$_{19}$ ceramics, one for the air-sintered specimen, the other for O$_2$ treated specimen. The resistivity of BaFe$_{12}$O$_{19}$ ceramic with O$_2$ treatment is around 8.5×10$^8$ Ω at 0.01Hz, while the one without oxygen treatment exhibits the resistivity at only 5.1×10$^5$ Ω, which is around 3 orders lower than that of O$_2$ annealed specimens. The huge increment of the ceramic's resistivity may be explained by the distinctively subduction of the density of oxygen vacancies and Fe$^{2+}$ oxided into Fe$^{3+}$ that relieves the current leakage raised from oxygen vacancies and electronic hopping between Fe$^{2+}$ and Fe$^{3+}$ ions.

### 3.3 Dielectric Relaxation Behavior of BaFe$_{12}$O$_{19}$ Capacitors

*Figure 3*A demonstrates a plot of the variation of dielectric constant of BaFe$_{12}$O$_{19}$ ceramic with temperature at the frequency of 100Hz. There appeared two dielectric anomaly peaks at the vicinity of Curie temperatures; one is a small hump peak (T$_d$) at 194°C and the other big one for the maximum dielectric constant peak (T$_m$) at 451°C. The first short peak is proposed to be the phase transition from ferroelectrics to anti-ferroelectrics, while the second main peak is supposed to be the phase transition from anti-ferroelectrics to para-electrics. This temperature-dependent dielectric performance is consistent with that of PbFe$_{12}$O$_{19}$ and SrFe$_{12}$O$_{19}$ ceramics. [26, 33]

*Figure 3*B shows the dielectric relaxing behavior of the BaFe$_{12}$O$_{19}$ ceramics, i. e. the temperature-dependent dielectric constant upon different frequencies. The first anti-ferroelectric phase transition peak (T$_d$) shifts to higher temperature side with frequency, exhibiting large sensitivity and diffuse performance of the dielectric response to frequency. When the frequency increases from 10 Hz to 30 Hz, the small T$_d$ peak moves from 166°C to 178°C, while the top dielectric constant decreases from 2256 to 1780. When frequency further increases from 50 Hz to 5 kHz, T$_d$ disperses from 190 to 252°C and the highest dielectric constant further drops from 1354 to 800. However, the second T$_m$ peak show much low sensitivity to frequency. The dielectric constant of BaFe$_{12}$O$_{19}$ ceramics changes to be negative at higher frequencies (f>500Hz) when the temperature passes through 396°C. [33]

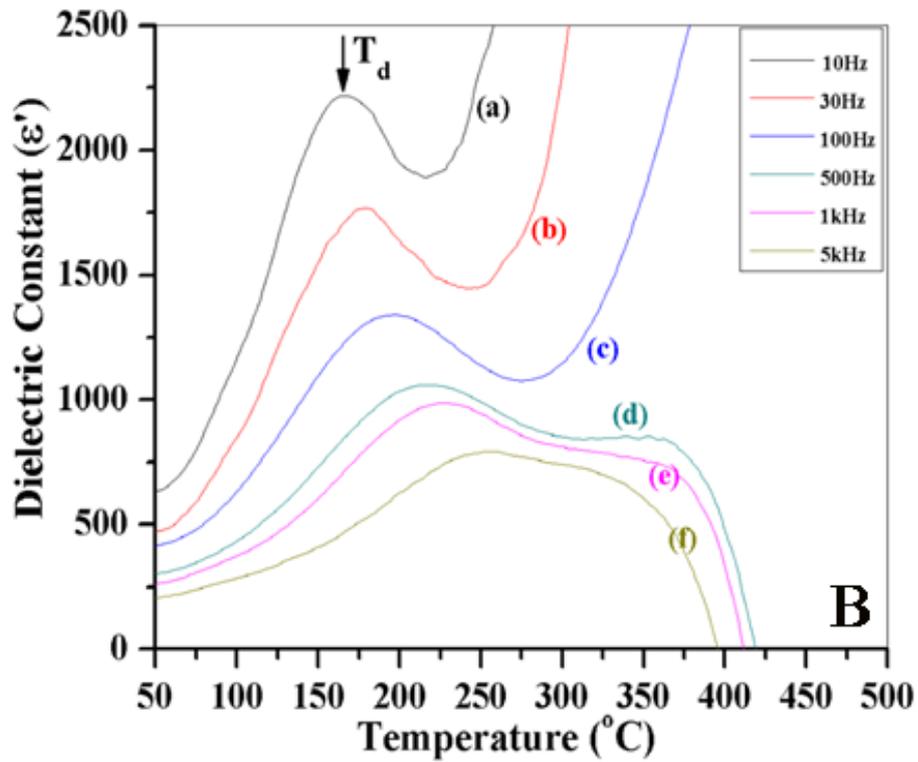
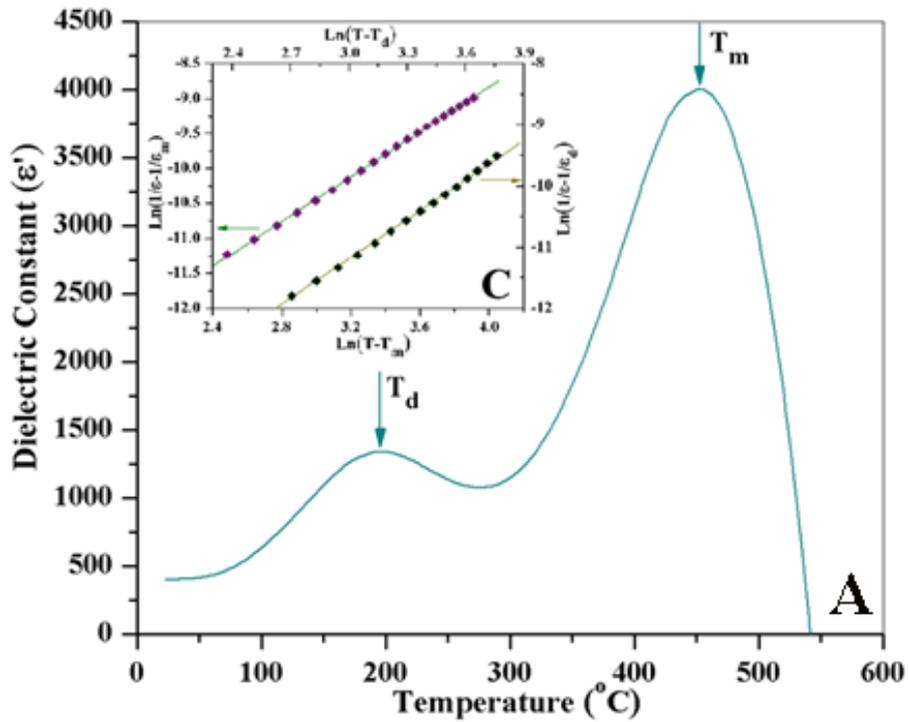

*Figure 3-(A): The temperature-dependent dielectric spectrum of BaFe$_{12}$O$_{19}$ ceramics at 100Hz. (B): The temperature-dependent dielectric spectra upon different frequencies with diffuse phase transitions (a) 10 Hz, (b) 30 Hz, (c) 100 Hz, (d) 500 Hz, (e) 1 kHz and (f) 5 kHz. (C): Curie-Weiss law verification- (a) the plot of log(1/$\varepsilon$-1/$\varepsilon_m$) - log(T-T$_m$) and (b) Log(1/$\varepsilon$-1/$\varepsilon_d$) - log(T-T$_d$) at 100 Hz for BaFe$_{12}$O$_{19}$ ceramics.*

For the relaxor ferroelectrics, the temperature-dependent dielectric constants in the vicinity of transition temperature should obey Curie Weiss law, which is expressed as the following formula [35]:

$$\frac{1}{\varepsilon} - \frac{1}{\varepsilon_m} = C(T - T_m)^\gamma$$

Where γ is the critical exponent, reflecting the transition diffuseness; $C$ is a constant. Dielectric constant $\varepsilon_m$ is corresponding to $T_m$. γ = 1 is for a normal ferroelectrics. When 1 < γ < 2, the material is called diffuse transition [36]. While γ > 2, the material is related to the anti-ferroelectric phase transition [26, 36].

*Figure 3*C shows the plots of Ln(1/ε-1/$\varepsilon_m$) and Ln(1/ε-1/$\varepsilon_d$) as functions of Ln(T-$T_m$) and Ln(T-$T_d$) at 100 Hz for $BaFe_{12}O_{19}$ ceramics, respectively. Linear fit to the experimental data matches the Curie-Weiss law very well. The fitting process produces two critical exponents, whose values are estimated to be $\gamma_m$ =2.173 and $\gamma_d$ =2.323. Both exponents are larger than 2, suggesting that $T_m$ and $T_d$ peaks are corresponding to the diffuse phase transitions being associated with antiferroelectricity as the intermediate phase.

### 3.4 *Ferroelectric properties of $BaFe_{12}O_{19}$ capacitors*

The previous work reported on the electric polarization (P-E) hysteresis loops for M-type barium hexaferrite ($BaFe_{12}O_{19}$) ceramics exhibited linear "banana" shape,[31] which provided some primitive evidence to reveal the doubtful ferroelectricity of $BaFe_{12}O_{19}$. Afterwards, several research groups published a few papers on the P-E hysteresis loops of $BaFe_{12}O_{19}$ ceramics with linear shapes, [29, 30, 32] trying to claim the ferroelectricity of $BaFe_{12}O_{19}$. However, those "banana" shaped P-E hysteresis loops without saturation were not powerful enough to prove the intrinsic ferroelectricity of $BaFe_{12}O_{19}$. They may mislead our views to the consequence of current leakage.

Usually, the "banana" shaped P-E loops are considered to be closely associated with current leakage, or even to be an output of parallelly linked capacitor - resistor circuit. Therefore, these P-E loops could not be used to be the experimental evidence to prove the ferroelectricity. Till nowadays, the ferroelectricity of $BaFe_{12}O_{19}$ ceramics still remains controversial and doubtful. Therefore, more powerful evidences are needed to convince the ferroelectricity of $BaFe_{12}O_{19}$. In order to achieve the saturated P-E loop, the current leakage should be greatly degraded. Under this consideration, we then annealed the $BaFe_{12}O_{19}$ ceramics in a tube furnace filled with pure $O_2$ at 700°C for different duration and step wises, so as to remove most portion of the oxygen vacancies and oxidizing $Fe^{2+}$ to $Fe^{3+}$. In this way, the current leakage would be substantially reduced. Then we may be able to watch out how $O_2$ annealing process would affect the concentration of charge carries and then the evolution of the P-E loop with the current leakage.

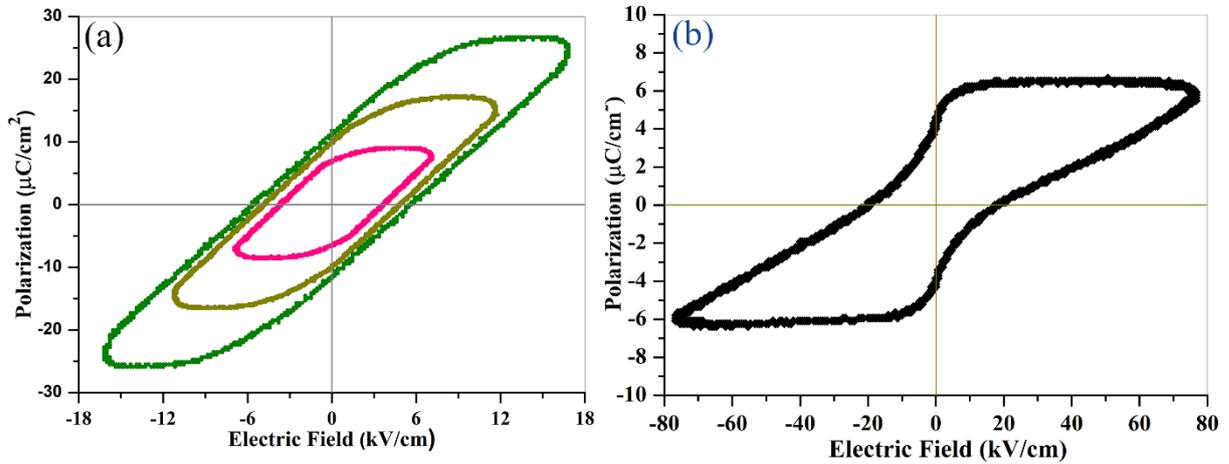

*Figure 4: The electric polarization hysteresis loop of $BaFe_{12}O_{19}$ ceramics, (a) being sintered at 1250°C in air only, (b) being sintered at 1250°C in air and subsequent annealing in oxygen for 3 hours once.*

We then made the ferroelectric measurement on three types of specimens, one is the as-sintered $BaFe_{12}O_{19}$ ceramic, which was subsequently annealed at 700°C in pure $O_2$ for 3 hours once and three times of 9 hours in 3 stepwise as the second and third types. Figure 4 shows the P-E hysteresis loops of the first two types of the specimens. Obviously *Figure 4*a shows a linear P-E hysteresis loop with a "banana" shape, indicating huge current leakage and leaving itself as a doubtful evidence for the ferroelectricity. After one stepwise oxygen annealing process for 3 hours, the P-E loop is getting better, exhibiting some non-linear feature in the P-E hysteresis loop with certain saturation. The heat treatment of $BaFe_{12}O_{19}$ ceramics in oxygen removes part of oxygen vacancies and transforms some $Fe^{2+}$ into $Fe^{3+}$, such that the concentration of charge carriers for current leakage was reduced somehow, resulting in better saturation of the P-E hysteresis loop.

We then extend the duration of oxygen annealing process of $BaFe_{12}O_{19}$ specimens to 9 hours at 700°C in 3 step wises, so as to further reduce the concentration of those charge carriers and greatly enhance the resistivity as we did in $PbFe_{12}O_{19}$[26] and $SrFe_{12}O_{19}$[33] ceramics before. It is supposed that longer duration of $O_2$ treatment would greatly degrade the concentration of defects of oxygen vacancies and $Fe^{2+}$ ions, the current leakage would then dramatically be reduced and thus saturated P-E loops could be achieved.

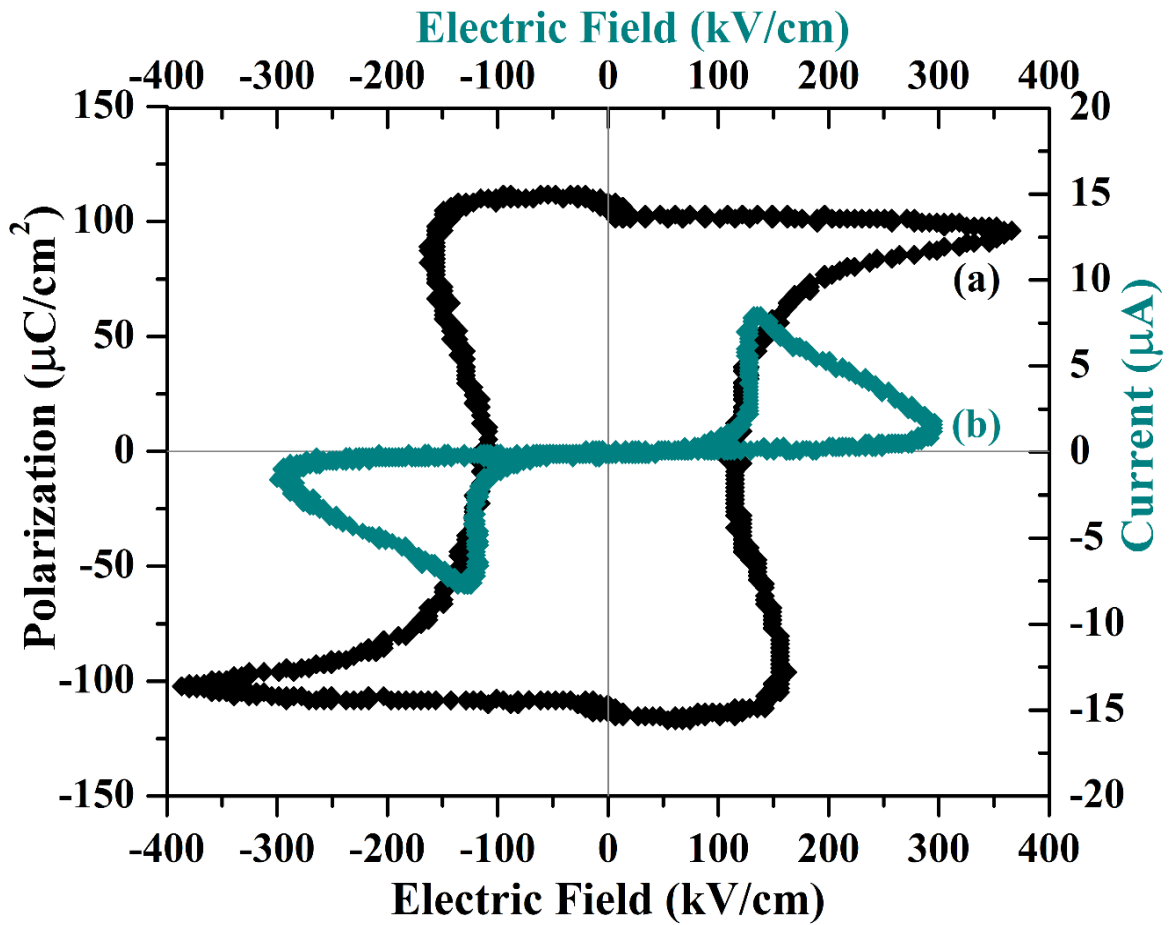

*Figure 5: (a) The fully saturated polarization hysteresis (P-E) loop of $BaFe_{12}O_{19}$ ceramic. (b) Plot of current versus voltage curve (I-V curve) for $BaFe_{12}O_{19}$ ceramic. The ceramic was sintered at 1250°C for 2 hours and subsequently heat-treated in $O_2$ for 9 hours in 3 stepwise.*

*Figure 5*(a) shows such a fully saturated ferroelectric polarization hysteresis (P-E) loop with a classic shape for the third type of $BaFe_{12}O_{19}$ ceramic, which was sintered at 1250°C for 2 hours and subsequently heat treated in pure $O_2$ at 700°C for a total duration of 9 hours in 3 stepwise. The ferroelectric measurement was carried out at room temperature (300K) upon the a frequency of 33Hz.

When the applied field decreases from the top saturated voltage, the polarization remains almost the same value without variation until the field switches to the opposite direction at negative coercive force (-360 kV/cm), then the polarization drops suddenly down to negative saturation value. The path of the alternative polarization is in agree with Landau's thermodynamic theory for ferroelectrics. The extrusive convex parts in the hysteresis loop at second and fourth quadrants probably imply the existence of a small fraction of antiferroelectric phases. It may be seen from *Figure 3*A that the phase transition from ferroelectrics to antiferroelectrics starts at around 50°C, which is very close to room temperature. Therefore, it's pretty normal that the room temperature P-E hysteresis loop of $BaFe_{12}O_{19}$ ceramics contains partial feature of anti-ferroelectric phase. The value of saturation holds almost the same line with remnant polarization reflects that most of the ceramic's polarization domains keep aligning themselves along with the direction of the applied electric field, while the dipoles in other domains aligning in other directions may counteract each other and make no contribution to the macroscopic polarization. The polarization field switches to reversal direction along with the applied electric voltage, resulting in the change of the polarization signs to be negative along with a reversal concave arc line (*Figure 5*) over the coercive field range (-108 kV/m to -382 kV/m). The coercive field of the $BaFe_{12}O_{19}$ ceramic was estimated to be 117kV/cm, while the remnant polarization was determined to be 108μC/cm$^2$ which is almost one order larger than the reported value at 11.6μC/cm$^2$ for the same specimen without oxygen annealing process [31]. Hence this step of $O_2$ treatment on BaFe12O19 ceramics can saturate the hysteresis loop and increase the remnant polarization value by eliminate oxygen vacancies and promote iron violence to alleviate the current leakage [26, 33]. Several P-E hysteresis loops from different $BaFe_{12}O_{19}$ ceramic specimens are supplied in the Supplementary Materials (Figure 1S) which aims to confirm the reliability and repetitiveness of the ferroelectric data.

*Figure 5*(b) shows a plot of current as a function of voltage (I-V curve) of $BaFe_{12}O_{19}$ ceramics. Two nonlinear I-V curves are assigned to the polarization switches along with the direction change in applied electric field. When the ferroelectric polarization switches at coercive field, there shall appear an momentarily up-rush or down-rush current flow from the instantaneous unanimous alignment of polarization, which results in the surface charges flowing instantaneously from one side to the other. The current stream switching happens synchronously with that of polarization charges at coercive field. Therefore, two reversal current peaks would appear in the I-V curve, as being shown in *Figure 5*(b). The two nonlinear I-V peaks clearly demonstrate the phenomenon of the polarization or domain switching, no linear components for current leakage shall match this feature. Non-ferroelectric materials or conventional dielectric medium could not be able to show up such non-linear I-V peaks with up or down rush momentary current flow, they would generate linear I-V lines instead. These two current peaks in the I-V curve resemble closely that of traditional perovskite compounds ($Pb(Zr_{0.4}Ti_{0.6})O_3$ and $LiNbO_3$) [37], may provide an additional fingerprint evidence for the ferroelectricity of $BaFe_{12}O_{19}$ ceramics. These two non-linear I-V peaks, together with the huge change of dielectric constants at the phase

transition temperature, would confirm that the hysteresis loop in *Figure 5*(a) was indeed induced by the ferroelectric polarization rather than by linear current leakage.

### 3.5 *Magnetic Properties of $BaFe_{12}O_{19}$ Compound*

The magnetic property was measured upon the $BaFe_{12}O_{19}$ powders by the Physical Property Measurement System (PPMS). The specimen was fabricated by the same process with the bulk ceramics. The magnetic hysteresis loop is demonstrated in *Figure 6*. The coercive field of the $BaFe_{12}O_{19}$ powders is around 2150 Oe and the remnant magnetic moment is 35 emu/g. These magnetic values are comparable to that of $PbFe_{12}O_{19}$ compound and is much stronger than $BiFeO_3$.[5] The coercive force was enhanced around 543 Oe [31], while the remnant magnetic moment remains almost the same in comparison with the specimen without $O_2$ treatment. The increment of the magnetic performance comes from the contribution of higher concentration of $Fe^{3+}$ ions, which provide one more electron spin than $Fe^{2+}$. Therefore, $O_2$ heat treatment would not only saturate the P-E hysteresis loop, but also improve the magnetic properties of $BaFe_{12}O_{19}$ through oxidizing $Fe^{2+}$ to $Fe^{3+}$.

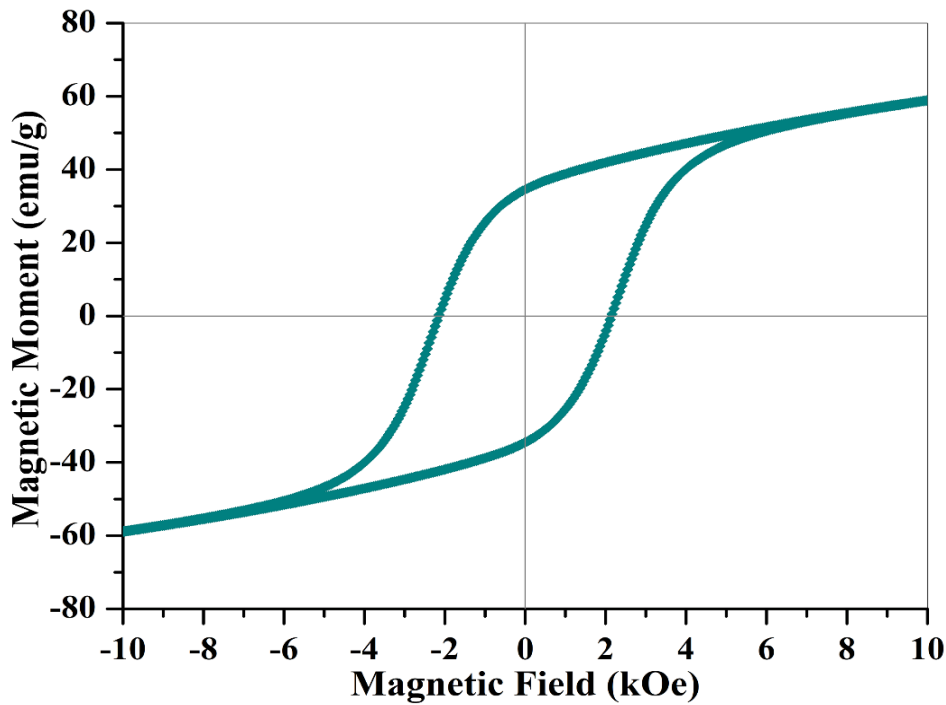

*Figure 6*: *Magnetic hysteresis loop of $BaFe_{12}O_{19}$ specimen, which was sintered at 1250°C for 2 h and subsequently annealed in $O_2$ for 9 hours with 3 steps wise.*

After the appearance of large ferroelectricity in $BiFeO_3$ thin film [5], the contradiction between conventional ferroelectricity and magnetism has caused a flurry of research intended to identify the new mechanisms of coexisting electric order and magnetic order [9]. It is one typical representative of perovskite- structure that bismuth ferrite: $Fe^{3+}$ with five unpaired elections in 3d orbit providing the magnetism and $Bi^{3+}$ with lone pair electrons from structural distortions providing the ferroelectricity.

In most of ferroelectric materials, the origin of the polarization is the move of nonmagnetic cations departing from the center of octahedron being surrounded by the neighboring anions. The relative displacement of anions and captions is not equal, leading to the noncentrosymmetricity and thus produces electric dipole moment. However, in conventional magnetic materials, the magnetic cations always locate rightly at the center of the neighboring anions [9], which requires centrosymmetric structure. Therefore, the electric polarization conflicts with magnetic moments in the symmetry of the structure. It can be understood that the difference lies in the competition consequence between energy-lowering covalent bond formation (larger for $d^0$ cations with their empty $d$ shells) and energy-raising electronic Coulomb repulsion (larger for cations with $d$ electrons) [9], which results in nonmagnetic and magnetic performance, respectively. That balance benefits off-centering of $d^0$ ions such as the $Ti^{4+}$ ions in barium titanate [9] or $Fe^{3+}$ in $FeO_6$ octahedron of $PbFe_{12}O_{19}$ [25, 26], which induces charge dipoles and thus trigs the prototypical polarization for ferroelectricity. In contrary, magnetic behavior appears in such candidates that have unpaired electron spins, such as the transition metals with partially filled $d$ orbits, which is again exactly the $Fe^{3+}$ in case of $BaFe_{12}O_{19}$. The off-center displacement of $Fe^{3+}$ in $FeO_6$ octahedron distorted the sub-lattice of $BaFe_{12}O_{19}$ and thus brings forth a local noncentrosymmetric structure distortion to lower symmetricity, which produces the electric dipole and then ferroelectricity. On the other hand, the unpaired electron spins within partially filled 3d orbits of these $Fe^{3+}$ ions produce magnetism. So $Fe^{3+}$ ions play both key roles on lowering the local symmetry at the scale of sub-lattice dimension to noncentrosymmetric structure as well as providing unpaired electron spins, the former one leads to the ferroelectricity and the latter one produces ferromagnetism. Gaining more $Fe^{3+}$ ions in $BaFe_{12}O_{19}$ through annealing it in $O_2$ atmosphere is highly desired to realize the multiferroics in M-type Barium Hexaferrite ($BaFe_{12}O_{19}$). The coexistence of large ferroelectricity and strong ferromagnetism in pure $BaFe_{12}O_{19}$ was thus achieved by merging off-centered $FeO_6$ octahedron and free spins in $Fe^{3+}$, which gives rise to the multiferroics in M-type barium hexaferrites ($BaFe_{12}O_{19}$).

### 3.6 Magnetoelectric Polarization in $BaFe_{12}O_{19}$ Ceramics

The magnetoelectric coupling effect was measured by applying a DC magnetic field upon the $BaFe_{12}O_{19}$ capacitors, the field was perpendicular to the surfaces of the pellet capacitors. The surfaces of the pellets were coated with silver electrode, which was linked to the two cables of Keysight 2450 source meter. The ME coupling voltage stimulated by the magnetic field was then output to the source meter. The output voltage from the source meter was set to be zero. In this way, we could ensure that the output voltage on the source meter is pure magnetoelectric coupling signals. The $BaFe_{12}O_{19}$ pellet ceramic was annealed in $O_2$ atmosphere as such the resistivity is over 500 MΩ, showing highly insulating features. Therefore, the measurement would not produce current leakage. In this way, we could be able to avoid the confusion of the current leakage with our ME signals.

*Figure 7* shows the alternative ME coupling voltage upon the magnetic field of 9mT and 1000 mT, respectively. Curve (a) is corresponding to the alternative ME coupling voltage upon an applied magnetic field of 9 mT. The coupling voltage appears in form of a cosine spin waves. The amplitude of the coupling voltage wave is around ±7 mV. Curve (b) is for the alternative ME coupling voltage upon an applied magnetic field of 1000 mT. The amplitude of the coupling voltage wave is around ±23 mV. There is an abrupt lift for the amplitude of coupling voltage wave when the applied magnetic field changes from 9 mT to 1000 mT. The variation of the amplitude is around 16mV.

This kind of coupling voltage wave arises from the magnetically induced spin current wave, which would then generate polarization in multiferroic materials. The H-induced ferroelectricity in our M-type barium hexaferrite was attributes to the spin-orbit coupling driven inverse Dzyaloshinskii-Moriya-type antisymmetric mechanism [38], where in the two nonequivalent canted FM spin pairs are coupled, non-collinearly interacting with each other, as it did in other ferrites of $SmFeO_3$ and $DyFeO_3$.[38] It appears that the exchange striction is responsible for the coupling polarization here in $BaFe_{12}O_{19}$ as well. Stronger magnetic field induced larger spin current wave or higher polarization voltage without any delay, thus causes the sudden rise of the coupling voltage when the magnetic field increases. This magnetoelectric coupling (ME) phenomenon[1, 39] produces a sizeable change of coupling voltage in response to the applied magnetic field from the alignment of the two electromagnets, which results in the appearance of spin current waves, as being captured by Keysight 2450 source meter and shown in *Figure 7*(c).

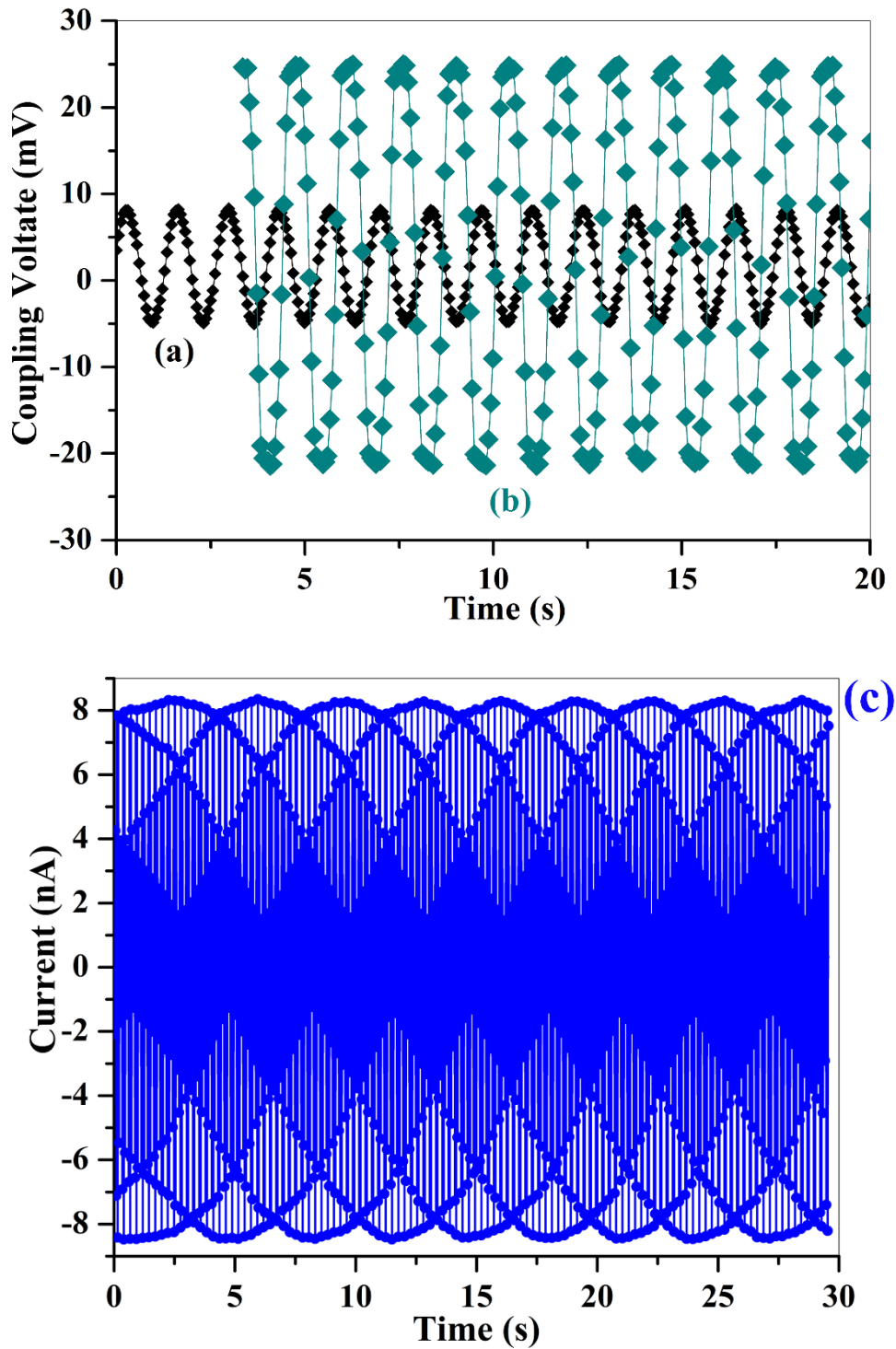

*Figure 7: The magnetoelectric coupling voltage waves of BaFe$_{12}$O$_{19}$ ceramics upon an applied magnetic field at (a) 9mT and (b) 1000mT; (c) the spin current waves of BaFe$_{12}$O$_{19}$ ceramics upon a magnetic field at 1000mT.*

The measured spin current wave is consistent with the theoretical calculation results [40]. Katsura et. al. did some theoretical work on spin current and magnetoelectric effects in noncollinear magnets, where they predicted that the spin current in the spiral magnets is

expressed in the following equation [40]:

$$P = \sum_j P_{j+1/2} \propto \sum_j \sin^2\beta \cos\alpha \sin(\gamma_{j+1} - \gamma_j);$$

$$i = \frac{dP}{dt} \propto -\sum_j \sin^2\beta \cos\alpha \cos(\gamma_{j+1} - \gamma_j).$$

Equation 3

Where, α is the angle of the spiral spin with z axis, β is the cone angle and γ is the angle of the spiral spin with *zx* plane; *j* and *j+1* are the atomic sites of transition metals ($Fe^{3+}$). The angle γ changes instantly with time while the angles α and β changes with the magnetic field, which induces different magnetic structures in $BaFe_{12}O_{19}$. The M-type barium hexaferrites would take a spiral spin structure at the magnetic field of 1000mT, which would induce large electric polarization as a function of $\sin^2\beta\cos\alpha$. Therefore, the spin current measurement is a function of $\cos(\gamma_{j+1}-\gamma_j)$, which is demonstrated as spin waves being shown in *Figure 7*(c).

## Conclusion

In summary, we are able to determine the ferroelectricity of M-type Barium Hexaferrite ($BaFe_{12}O_{19}$) by several powerful evidences: the most important fully saturated P-E hysteresis loos, the two nonlinear current peaks in I–V curve, and the giant change of dielectric constant in the vicinity of Curie temperatures. These combined experimental results verify the ferroelectricity of magnetic $BaFe_{12}O_{19}$ compound. The "banana" shaped P-E hysteresis loops in $BaFe_{12}O_{19}$ ceramics without $O_2$ heat treatment was also presented to make the comparison with the one being carefully annealed in $O_2$ atmosphere. The classic shape of the P-E hysteresis loops with full saturation in carefully prepared specimen may provide much more convincing evidence to prove the ferroelectricity of $BaFe_{12}O_{19}$ ceramics. The remnant polarization of the $BaFe_{12}O_{19}$ ceramics is around 108 $\mu C/cm^2$, while the coercive field is estimated to be 117kV/cm. Large magnetic hysteresis loop was also observed, exhibiting strong ferromagnetism in $BaFe_{12}O_{19}$. $Fe^{3+}$ ions are proposed to play both key roles on lowering the local symmetry at the scale of sub-lattice dimension to noncentrosymmetric structure as well as providing unpaired electron spins, the former one leads to the ferroelectricity and the latter one produces ferromagnetism. Heat treatment of the $BaFe_{12}O_{19}$ in $O_2$ atmosphere would gain more $Fe^{3+}$ ions and lead to much less oxygen vacancies, which are benefit to achieve both good ferroelectricity and strong ferromagnetism. Low B-field induced polarization and spin current waves have been captured from the generating ME coupling voltage in $BaFe_{12}O_{19}$ ceramics.

## Acknowledgement:

The authors acknowledge the financial support from the National Natural Science Foundation of China under grant No. 11774276.